\documentclass[]{tPHM2e}
\usepackage{graphicx}
\usepackage{wrapfig}

\begin{document}

\markboth{T. Yankova {\em et al.}}{}


\title{Crystals for neutron scattering studies of quantum magnetism}

\author{
T. Yankova$^\ast$,\thanks{$^\ast$Permanent address: Chemistry
Department, Moscow State University, Moscow, Russia.} D. H\"uvonen,
S. M\"uhlbauer, D. Schmidiger,\\ E. Wulf, S. Zhao and A.
Zheludev$^\dag$\thanks{$^\dag$ Corresponding author. Email: zhelud@ethz.ch.}\\
\vspace{6pt}  { \em Neutron Scattering and Magnetism Group,
Institute for Solid State Physics, ETH Z\"urich, Switzerland.}\\
T. Hong, V. O. Garlea, R.~Custelcean and G. Ehlers\\
{\em  Oak Ridge National Laboratory, Oak Ridge, TN, USA.}\\
\vspace{6pt}\received{\today}}

\maketitle

\begin{abstract}
We review a strategy for targeted synthesis of large single crystal
samples of prototype quantum magnets for inelastic neutron
scattering experiments. Four case studies of organic copper
halogenide $S=1/2$ systems are presented. They are meant to
illustrate that exciting experimental results pertaining to
forefront many-body quantum physics can be obtained on samples grown
using very simple techniques, standard laboratory equipment, and
almost no experience in in advanced crystal growth techniques.
\bigskip
\end{abstract}

\bigskip

\section{Introduction}
For the past 50 years, neutron spectroscopy has been one of the most
powerful tools for the study of local and collective excitations in
condensed matter systems \cite{Brockhouse1994}. Virtually all that
we know of phonons and magnons comes from such experiments. Most
other technique are not momentum resolved and provide only very
limited and indirect information. Other methods are restricted to
looking at the charge sector, and are not suited for the study of
magnetism and nucleus dynamics. Still others can only probe surface
properties.

Arguably, the main limitation of neutron spectroscopy is the need
for very large single crystal samples, typically of a few gram or
even tens of grams. Until roughly the mid-90s, this created a very
interesting synergy between crystal growers and expert neutron
scatterers. The complexity of neutron instrumentation and data
analysis, and the fact that neutrons are only produced at dedicated
large-scale installations, ensured that these were two distinct
groups of scientists. They were even separated geographically: the
former were typically at universities, while the latter concentrated
at the facilities.

In the past fifteen years the situation has changed drastically.
Many neutron techniques are now mature, well tested and fully
optimized. Instruments and data analysis software have evolved to be
accessible to the non-expert. Most importantly, most neutron
scattering facilities such as Institut Laue Langevin, ISIS and the
Spallation Neutron Source now run extensive programs to support
external users. Crystal growers nowadays can run their own neutron
experiments, while the neutron scattering experts serve them as
hosts and local contacts. Of course, not all experiments can be
carried out in this mode by non-experts, but {\em many} can.

What is a ``professional'' neutron scatterer to do, to maintain a
competitive research program of his or her own? A winning strategy
is to grow his or her own samples. Of course, not all materials can
be grown into large single crystals by a non-expert, but {\em many}
can. This paper is a case study. We show how a team of hard-core
neutron scatterers working in the rather esoteric field of quantum
magnetism can greatly benefit from investing just a little of their
time in crystal growth.

\subsection{Challenges in quantum magnetism} ``Quantum magnets'' are
defined as magnetic materials in which long-range magnetic order is
destroyed even at zero temperature by anomalously strong quantum
spin fluctuations. Many low-dimensional and geometrically frustrated
antiferromagnets (AFs) fall into this category. Since there is no
long range magnetic order, {\em all} the information on the physics
of these materials is contained in excitations. Examples include the
AF Heisenberg spin chains \cite{Bethe31} (including so-called
Haldane chains \cite{Haldane1983}), spin ladders \cite{Rice1993},
bond-alternationg chains \cite{Uhrig1996}, and more complex
frustrated geometries \cite{White1996}.

Neutron spectroscopy, being able to probe spin excitations at
non-zero momenta, played a particularly important role in
understanding the exotic and complicated many body quantum mechanics
on these materials. Initially, most studies were performed on
transition metal oxide systems. Among the examples are the Haldane
spin chain compound Y$_2$BaNiO$_5$, the spin chain system SrCuO$_2$
\cite{Zaliznyak1999} and the exotic spin ladder
La$_4$Sr$_{10}$Cu$_{24}$O$_{41}$\cite{Notbohm2007}. Unfortunately,
the rather large energy scale of magnetic interactions in such
materials (typically 10--1000~meV) prohibits the study of
finite-temperature effects ($\kappa_\mathrm{B}T\lesssim 30$~meV) and
the extremely interesting new physics \cite{Giamarchi2008} that
emerges in applied external magnetic fields
($g\mu_\mathrm{B}H\lesssim 2$~meV). Recent breakthroughs were
therefore achieved in studies of organic quantum magnets,
particularly in transition metal complexes. Among these compounds,
for neutron experiments, one finds excellent one-dimensional
\cite{Regnault1994,Stone2003,Kenzelmann2004,Masuda2006,Ruegg2008}
and two-dimensional \cite{Stone2001,Stone2006,Hong2011} spin
networks with varying degrees of geometric frustration
\cite{Stone2001,Garlea2009} and typical magnetic energy scales of
1~meV. These energies correspond to temperatures
\cite{Renard1988,Zaliznyak1994,Zheludev1996,Zheludev2008} and
magnetic fields
\cite{Zheludev2002,Hagiwara2005,Zapf2006,Garlea2007,Thielmann2009}
that can be realized in neutron experiments. They are also perfectly
suited for spectroscopic studies with low-energy neutrons
($\lambda\sim 5$~\AA). Moreover, the use of cold neutrons often
helps avoid background due to phonon scattering, and for a number of
technical reasons allows a cleaner measurement of the spin
excitations. In addition to temperature- and field-dependent studies
of quantum magnets, there is a growing interest in the effects of
disorder. For these applications too, transition metal organic
complexes are excellent models: chemical disorder is easy to
introduce on either the magnetic
\cite{Glarum91,Granroth1998,Uchiyama1999} or non-magnetic
\cite{Tippie1981,Manaka2001,Manaka2008,Hong2010} sites.

From the practical point of view of the neutron scatterer who
aspires to synthesize his or her own samples, transition metal
organic complexes often have the advantage of being easy to grow. In
this work we shall specifically focus on organic halogenides, for
which very large and nicely faceted single crystals can be obtained
in many cases. The growth is typically done in solution, and
requires no special equipment beyond that found in a typical wet
chemistry laboratory. Even physics students with minimal experience
in material science or chemistry can have the privilege of growing
(and modifying, and controlling) their own samples for neutron
experiments.

\subsection{Methods}
The peculiarities of neutron scattering impose specific requirements
on the samples used. As mentioned, large single crystals are a
necessity. Being large in volume, they should be rather compact. The
typical neutron beam cross section is only about 2~cm, and the
sample should be fully immersed. Long needles or thin platelets are
not particularly useful. Fortunately, for spectroscopy experiments,
the quality of the crystals is less critical than one may expect. A
mosaic spread of $2^\circ$ or even larger is acceptable. Several
grains aligned within this tolerance can be treated as a ``single''
crystal. Diffraction is more demanding, but even there a symmetric
mosaic spread of $0.5^\circ$ can often be tolerated. The neutron
scattering and absorbtion cross sections are strongly
isotope-dependent \cite{Squiresbook}. Hydrogen, for example, is
pernicious for neutron experiments due to its enormous incoherent
cross section. Incoherent scattering contributes to background and
removes neutrons from the incident beam. The neutron penetration
depth in a proton-containing sample is typically only a few
millimeters. Thus, in most cases, hydrogen-bearing samples for
neutron experiments must be deuterated. As will be illustrated in
one example below, partial deuteration is of not much use, and one
typically tries to substitute at least 95\% of all hydrogen sites
with deuterium (more is better). There are a few other ``taboo''
isotopes besides hydrogen \cite{Sears1992}. Among them are $^{10}$B,
natural Gd and natural Cd, due to their huge absorbtion cross
sections for thermal neutrons.

\begin{wrapfigure}{l}{0.5\textwidth}
 \subfigure[]{\includegraphics[height=0.38\textwidth]{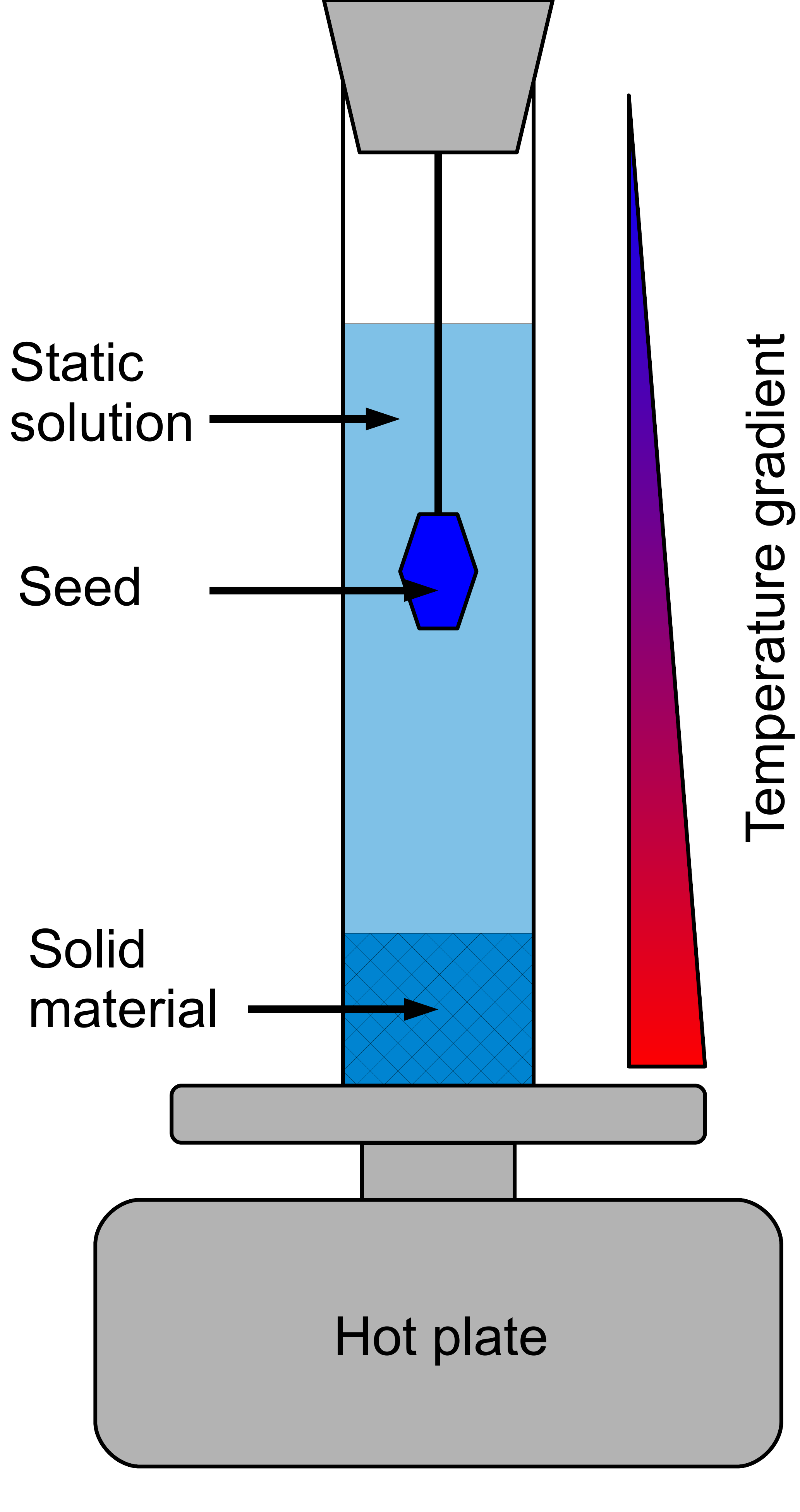}}
 \subfigure[]{\includegraphics[height=0.38\textwidth]{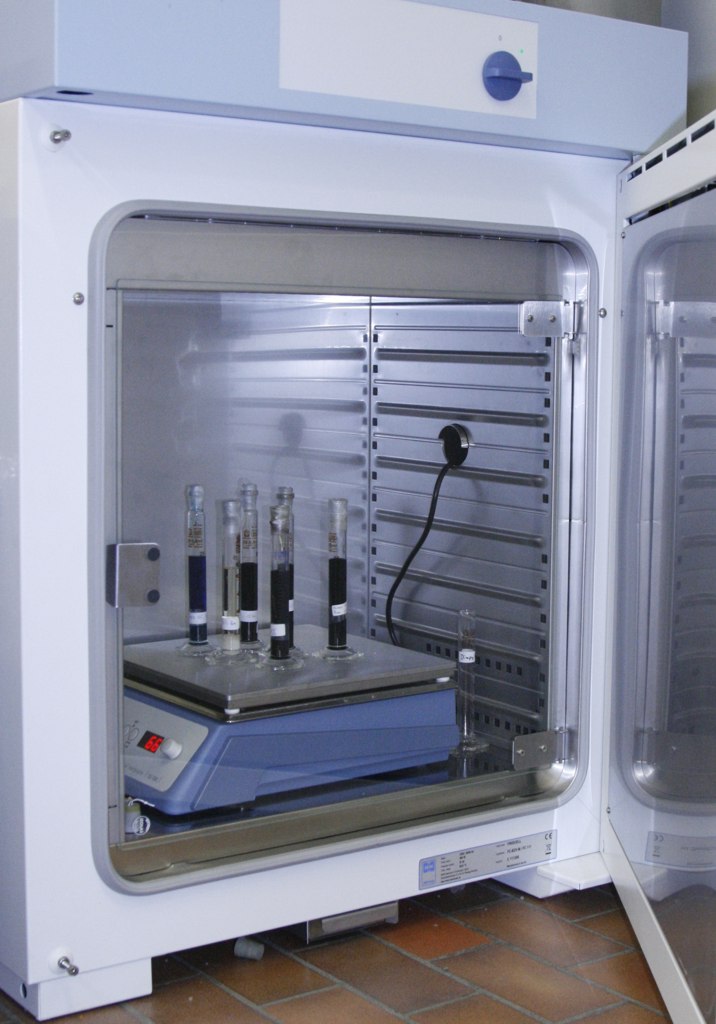}}
\caption{(Temperature gradient growth of single crystals: (a)
Schematic of the setup. (b) Refrigerated environmental chamber with
several crystallization cylinders on a hotplate.}
\end{wrapfigure}

By far, the easiest way to produce neutron-specific crystals is by
solution growth. Since it occurs at conditions close to equilibrium,
high-quality large crystals can be obtained. Low temperature
solution growth is particularly useful and versatile
\cite{crystalhandbook}. It is cheap, simple and relatively safe.
This technique also has the obvious advantage for crystals that are
unstable at high temperatures. Such are the transition metal organic
complexes that are a focus of this work. Several methods of
low-temperature solution growth can be classified: (i) slow cooling,
(ii) slow evaporation, (iii) the temperature gradient method, and
(iv) the chemical/gel method \cite{crystalhandbook}. Only the slow
evaporation and temperature gradient methods are discussed here.

\subsubsection{Slow evaporation}
Slow evaporation is useful for materials with a low temperature
coefficient of solubility. The growth occurs at a fixed temperature.
Due to the higher stability one can, in principle, expect larger and
better crystals than those obtained with slow cooling. The slow
evaporation method is very straightforward to realize. The key
requirement is high temperature stability. In our laboratory at ETH
Zurich it is achieved in commercial environmental chambers such as
the Friocell models by Medcenter Einrichtung GmbH. The
crystallization beakers are set up inside airtight containers (we
use standard glass dessicators), which are in turned placed in the
environmental chambers.

The removal of solvent vapors is performed in a controlled way, by
flowing nitrogen gas through the dessicators. The gas flow rate is
controlled by automated metering valves. The nitrogen and solvent
vapor exhaust is released into an acid hood. One of the many
advantages of this setup is the use of dry nitrogen gas that
prevents environmental water vapor from entering the growth
solution. Besides disrupting the growth balance, such water brings
in hydrogen ions. The latter can exchange with deuterium ions of the
growing crystal, which, for neutron scattering purposes, is to be
avoided. Some solvents are not particularly hygroscopic, and many
crystals will grow without a highly stable environment. In these
cases one can simply leave the growth beaker under an acid hood and
allow evaporation to take its course in an uncontrolled manner.

To obtain large crystals, seeds are first precipitated by slow
evaporation of a saturated solution. The best seeds are then
suspended on PTFE threads in a solution that is filtered and
saturated at the evaporation temperature. In our experiments the
latter varies between 20$^\circ$C to 60$^\circ$C, depending on the
substance to be grown and the solvent used. The crystal growth rate
and quality are controlled by temperature, nitrogen flow and
aperture of the growth beaker. Optimal growth parameters are
determined by trial and error, and luck is always an important
ingredient.

\subsubsection{Temperature gradient}
The main disadvantage of evaporative growth for our applications is
the loss of solvent. Deuterated organic solvents can be extremely
expensive and are to be conserved. Our attempts to recover solvent
vapor from the nitrogen exhaust using a Peltier-cooled condenser
were only partially successful. A totally different approach, namely
the temperature gradient method, is in many cases a better
alternative. In this technique, a thermal gradient is maintained
across the growth solution. Assuming that solubility increases with
temperature, an excess of the solute is placed in the hotter region.
The growth compound is transported through the solution to a cooler
region. Here the solution becomes over-saturated and crystals are
precipitated. The transport is usually due to thermal convection,
but can also be diffusive.

In our setup, we use volumetrical cylinders of volume 25~ml or
50~ml. As a first step, powder and seeds of the target compound are
obtained by slow evaporation. Some solid target material is placed
on the bottom of the growth cylinder and topped off with 25--50~ml
of the saturated solution. The seed is suspended in the upper region
of the cylinder, which is then sealed. A temperature gradient is
achieved by heating of the cylinder bottom on a hotplate
(Fig.~\ref{fig:envchamber}a). For stability, the hotplate is placed
into a refrigerated environmental chamber. Active refrigeration is
necessary to drain the heat released by the hotplate and ensure a
relatively low temperature for upper part of the cylinder. Typical
chamber temperatures are 30$^\circ$C, with hotplate temperatures
varying between 50$^\circ$C to 160$^\circ$C, depending on the
substances involved. Such a setup is illustrated in
Fig.~\ref{fig:envchamber}b.  Besides totally eliminating the loss of
solvent, this method provides isolation from any detrimental
environmental factors.

\subsubsection{Introducing disorder}
Among the issues that we address in the present study is the effect
of magnetic bond strength disorder on the properties and excitation
spectra of quantum spin systems. In our experiments on organic
copper halogenides, disorder is introduced by a partial chemical
substitution on the halogen sites. Crystals of such materials are
prepared by adding the corresponding reagents to the growth
solution. For example, instead of copper (II) chloride we use the
same molar amount of a mixture of copper (II) chloride and bromide.
The method works particularly well for partial substitution of Cl by
Br and {\em vice versa}, thanks to the similarity of the
corresponding ionic radii. The crystal growth procedures are the
same as for disorder-free parent compounds.

\begin{wrapfigure}{l}{0.4\textwidth}
\includegraphics[width=0.38\textwidth]{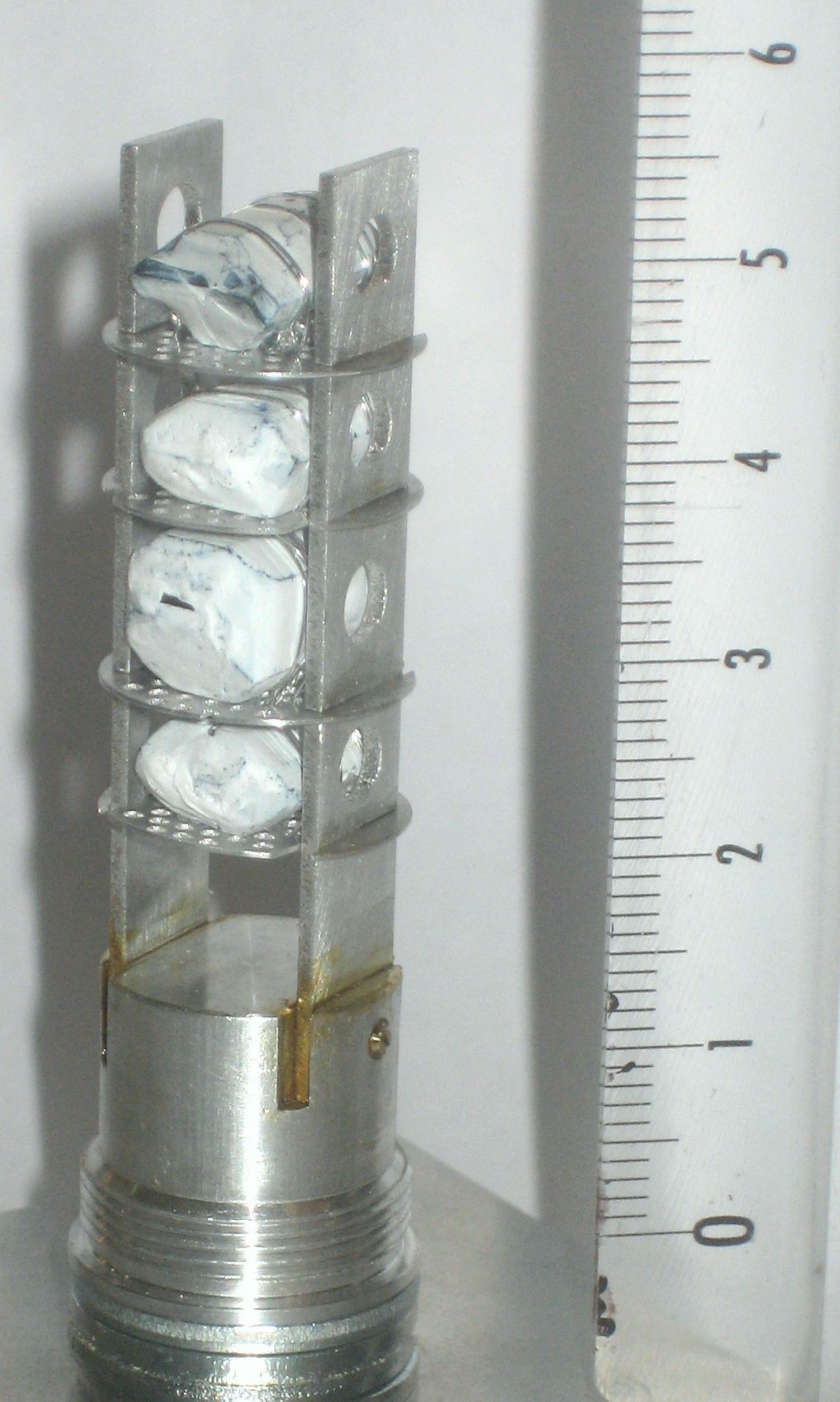}
\caption{Four deuterated crystals wrapped in PTFE film and wired
down on an aluminium sample holder for a total mosaic spread of
about 1$^\circ$. The use of perforated aluminium reduces the amount
of material in the beam. A small amount of glue visible near the
base of the holder is permissible, as it remains outside the neutron
beam. In this case, a stringent constraint on sample diameter is
imposed by the used of a $^3$He-$^4$He dilution refrigerator  to
achieve $T<100$~mK required for some experiments.
}\label{DIMPYsample}
\end{wrapfigure}

Typically, substitution at concentrations of up to 10 at.\% is
relatively easy to achieve. Crystal growth at higher concentration
is often inhibited. We speculate that this is due to large lattice
strains that emerge when at high concentrations two or more ions of
the ``wrong'' radius are forced on adjacent sites. In general, the
higher the substitution level, the lower the growth rate and the
lower the quality of the resulting crystals. In some cases, the
introduction of foreign halogen ions leads to the formation of
entirely different and typically useless (from the physics point of
view) crystallographic phases. It can also happen that only one
stoichiometric material will grow even from a mixture containing
different halogen species.

The nominal concentration of halogen-site substitute in solution can
be different from that in the resulting crystal. Fortunately, our
organic systems involve mostly light elements. As a result, the
contrast between Cl and Br is sufficient to determine the
concentrations with X-ray diffraction. Such studies are performed
using a  small-molecule single crystal diffractometer.  In our ETH
Zurich laboratory we employ the Bruker APEX II system. Often, the Br
or Cl content can be determined to within a fraction of a percent.
For a cross-check we also use micro-elemental analysis (Schoeniger
method), but that tends to be less precise for small concentrations.
A common concern is whether halogen concentrations remain uniform
even in very large crystals. To within the experimental error of our
chemical analysis, this appears to be the case for all materials
discussed here. In some cases, this conclusion can also be drawn
from the sharpness of magnetic phase transitions observed with
neutron diffraction.

\subsubsection{Preparing the crystals for experiment}

A few words need to be said about mounting the single crystals for
neutron scattering experiments. First, they need to be aligned in a
particular orientation to within a couple of degrees. Having nicely
faceted crystals that are often the result of solution growth, helps
with this task. The alignment can be done with X-ray Laue
diffraction or using a single-crystal diffractometer. Even if X-rays
are used, the final test is to be done with neutrons, since only
neutrons can penetrate the bulk of the crystal and reveal hidden
misaligned grains. Very often several separate crystals can be
co-aligned to produce a larger volume ``supersample''.  The sample
mounts for neutron experiments are typically made of aluminium or
copper, which are largely transparent for neutrons and have good
thermal conductivity. Even so, scattering from either metal can
contribute to the background, and Cu is slightly absorbing.
Therefore, as little material as possible is to be used. To avoid
introducing hydrogen in the beam, no conventional glue can ever be
applied! Instead, the crystals are attached with thin aluminium wire
or PTFE film (plumbing tape). PTFE is hydrogen-free and produces
surprisingly little incoherent neutron background.

Assembling and aligning such samples may take hours or even days of
work in the neutron beam (triclinic crystals are the hardest to deal
with). This implies exposing the crystals to the atmosphere. Sample
stability and hygroscopicity become serious issues. Some of our
organic copper (II) halogenides are unstable, being hygroscopic or
losing solvent that is part of their crystal structure to
evaporation. In addition, all of them are corrosive materials and
may react with the sample holders. Several strategies can be used to
protect the crystals. Tightly wrapping them in PTFE helps, as does
dipping crystals into PFPE lubricants such as Fomblin. An example of
four co-aligned crystals ready for neutron experiments is shown in
Fig.~\ref{DIMPYsample}.

\section{Case studies}

\subsection{A strong-leg spin ladder}
The antiferromagnetic $S=1/2$ Heisenberg two leg ladder is probably
the most famous model in quantum magnetism. It exhibits a gapped,
spin-liquid ground state with sharp triplet excitations. It can be
intuitively understood in two different coupling limits. A {\it
strong-rung} ladder is essentially a system of weakly coupled
dimers. In contrast, the more interesting {\it strong-leg} ladder is
described as weakly interacting quantum spin chains. Until the
discovery of the {\it strong-rung} spin ladder material
(C$_5$H$_{12}$N)$_2$CuBr$_4$ (Hpip) \cite{Patyal:90}, most known
ladders were cuprates with large intrinsic energy scales
\cite{Dagotto:99}. The low energy scales of organometallic Hpip for
the first time opened experimental access the entire magnetic phase
diagram  \cite{Ruegg2008-2,Thielmann2009,Thielmann2009PRB}, but
obviously only in the strong-rung limit.
\begin{figure}
\begin{center}
\includegraphics[width=\textwidth]{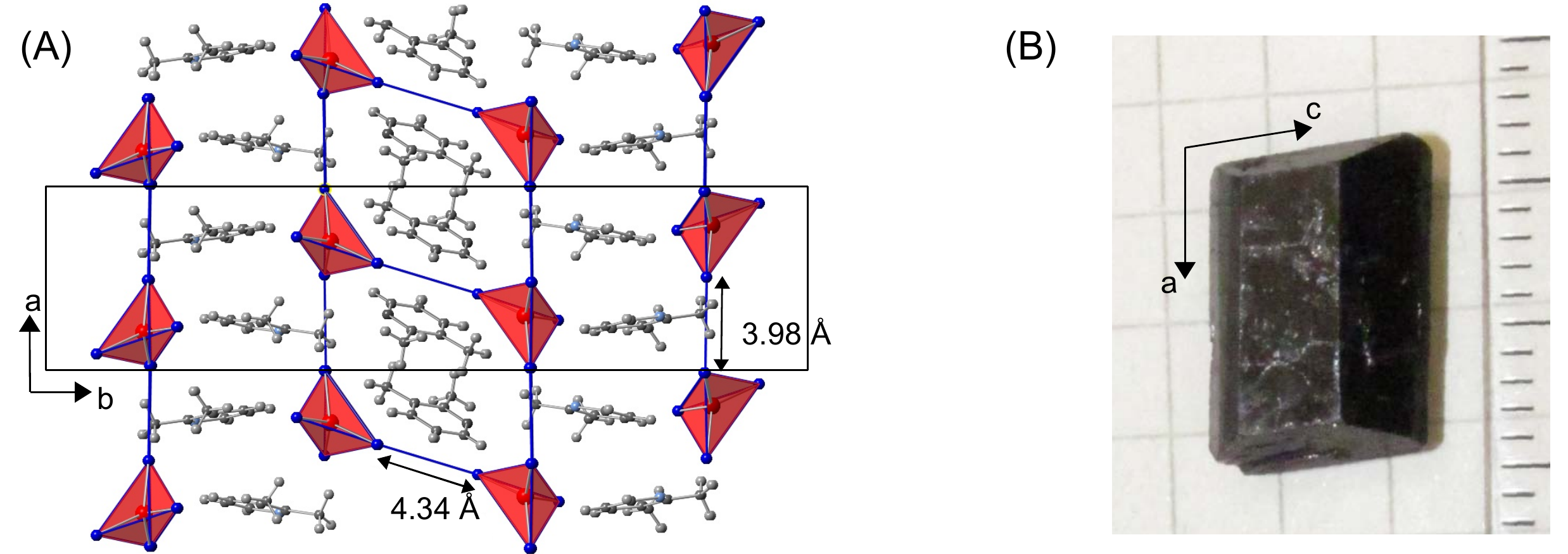}
\caption{\label{DIMPYstructure}(A) Crystallographic structure of
DIMPY seen along the $c$-axis: The Cu$^{2+}$-ions are colored in
red, the Br$^{-}$ions in blue. The CuBr$_4^{2-}$ tetrahedra are
shaded in red. The organic spacer  molecule is shown in gray. The
blue lines indicate the spin ladder structure. (B) Fully deuterated
single crystal of DIMPY with a size of $\approx 11 \times 8\times
5\,\,{\rm mm^3}$.}
\end{center}

\end{figure}

The first realization of a {\it strong-leg organometallic}
Heisenberg spin ladder is the compound
Bis(2,3-dimethylpyridinium)-Tetrabromocuprate (DIMPY)
\cite{Shapiro:2007, Somoza}. DIMPY crystallizes in the monoclinic
space group $P2_1/n$ with lattice parameters $a=7.504\,\,$\AA,
$b=31.61\,\,$\AA, $c=8.202\,\,{\rm \AA~}$ and $\beta=98.97^{\circ}$.
The magnetic behavior is governed by the Cu$^{2+}$-ions in a ladder
structure composed of CuBr$_4^{2-}$ tetrahedra. The Cu$^{2+}$-ions
are interacting via nearest neighbor Cu-Br$\cdot\cdot$Br-Cu AF
superexchange. The unit cell and the spin ladder structure is shown
in Fig. \ref{DIMPYstructure}(A). Measurements of susceptibility
\cite{Shapiro:2007} and specific heat \cite{Hong2010} are consistent
with a spin ladder in the {\it strong-leg} regime. The spin
excitation spectrum was studies by inelastic neutron scattering on a
partially (67\%) deuterated single crystal sample and confirmed a
spin gap of $\Delta= 0.32(2)\,\,\rm{meV}$ \cite{Hong2010}. However,
due to the strong incoherent scattering from hydrogen, the magnons
could be only observed at large wavelengths, close to the
antiferromagnetic zone center. It became clear that no further
progress can be made without {\it fully deuterated} single crystal
samples.

Such high quality fully deuterated (98 at.\% D) single crystals of
DIMPY were successfully grown in our laboratory at ETHZ using the
temperature gradient method. Due to the high costs, the growth
recipe has been optimized to use as little deuterated reagents as
possible: $0.03356\,\,\rm{mole}$ ($3.75\,\,\rm{g}$) of deuterated
2,3-Lutidine (C$_7$D$_9$N), $0.03356\,\,\rm{mole}$ of
$4.5\,\,\rm{M}$ deuterobromic acid (DBr) and $0.01668\,\,\rm{mole}$
($3.727\,\,\rm{g}$) of high purity water free CuBr$_2$ are
consecutively dissolved in $30\,\,\rm{ml}$ of ethanol-d$_6$ while
stirring at $50^{\circ}\,\rm{C}$. A dark violet solution is
obtained. After stirring for 24 hours in a sealed container, the
solution is slowly evaporated and $10\,\,\rm{g}$ of dry DIMPY powder
is harvested. $7\,\,\rm{g}$ of the powder is redissolved in a
$25\,\,\rm{ml}$ cylinder with $22\,\,\rm{ml}$ ethanol-d$_6$. The
result is a saturated solution with $\approx 3\,\,\rm{g}$ of
undissolved DIMPY at the bottom of the cylinder. The growth is
performed with the temperature gradient method at a base temperature
of $65^{\circ}\,\rm{C}$ and ambient temperature of
$30^{\circ}\,\rm{C}$. The seeds were separately obtained by the slow
diffusion method. After two months, single crystals with a mass of
$1\,\,\rm{g}-1.5\,\,\rm{g}$ are harvested. After refilling DIMPY
powder to maintain a saturated solution the growth procedure can be
repeated several times. The crystals show a dark violet to black
color and clean facets. Fig. \ref{DIMPYstructure}(B) depicts a
typical single crystal with a mass of $\approx 1\,\,\rm{g}$. The
crystalline structure and quality were confirmed with single crystal
X-ray and neutron diffraction, revealing a mosaic spread of
0.45$^{\circ}$ FWHM and no signs of parasitic phases.

\begin{figure}
\begin{center}
\includegraphics[width=\textwidth]{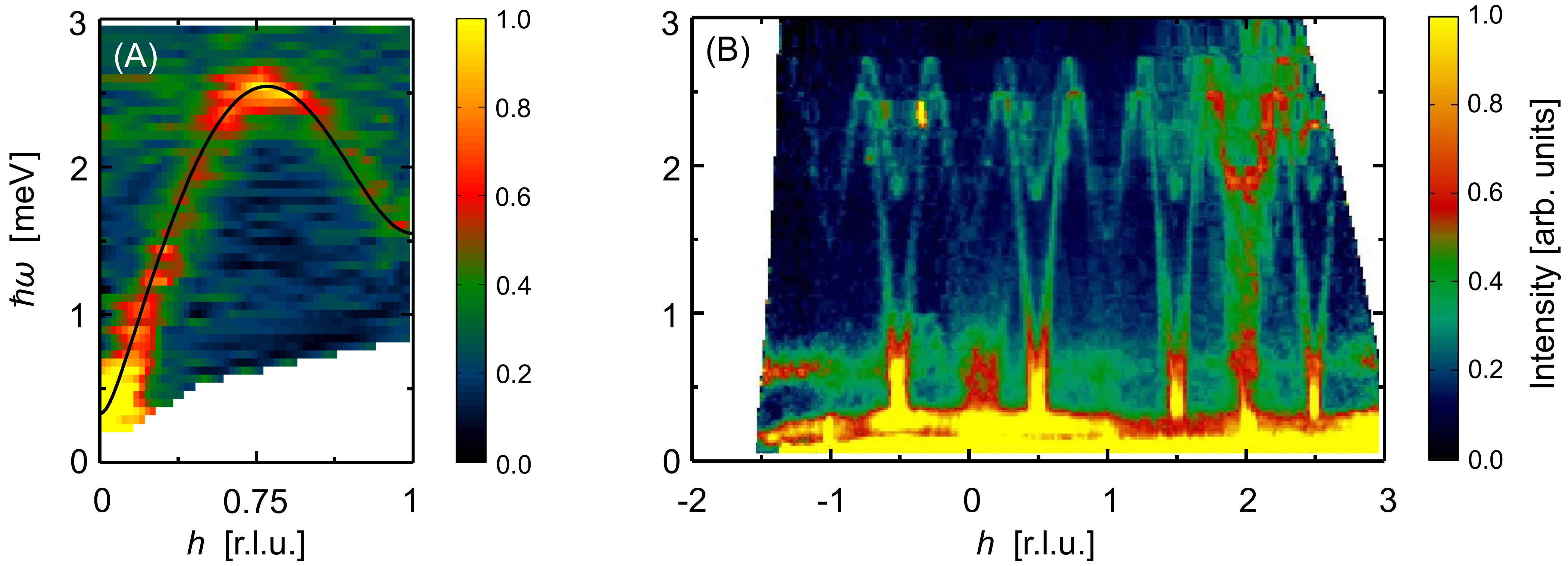}
\caption{\label{DIMPYdispersion}(A) False color map of the spin
excitation spectrum in DIMPY composed of individual constant-q scans
measured along the line ${\bf q}=(h,0,0.17-1.44\cdot h)$ for
$0.5<h<1$, $T=50\,\,{\rm mK}$. The data are from
\cite{Schmidiger:11}. (B) False color projection of the spectrum
measured on in time-of-flight mode at $T=2\,\,{\rm K}$
\cite{Schmidiger:11b}.}
\end{center}

\end{figure}

Experiments on these highly deuterated samples, some still ongoing,
are providing new insight on the physics of quantum spin ladders. We
are utilizing four co-aligned crystals with a total mass of
$3.7\,\,\rm{g}$ (Fig.~\ref{DIMPYsample}). The initial data were
collected on the triple axis spectrometer TASP at Paul Scherrer
Institut, Villigen, Switzerland. The spectrum measured at
$T=50\,\,{\rm mK}$ is shown in Fig.~\ref{DIMPYdispersion}(A). It
reveals that the single-magnon excitations remain sharp throughout
the entire Brillouin zone. This confirms a peculiar symmetry of the
ladder spin Hamiltonian which is invariant under leg permutation
\cite{Schmidiger:11}. In a previously studied asymmetric ladder the
magnons become unstable at certain wave vectors\cite{Masuda2006}.

The most recent experiments were carried out at the Cold Neutron
Chopper Spectrometer CNCS \cite{Ehlers2011} at Oak Ridge National
Laboratory and revealed stunning new details \cite{Schmidiger:11b}.
A projection of the 4-dimensional data set onto the energy- leg
momentum axes recorded during a single day on of measurement is
shown in Fig. \ref{DIMPYdispersion}(B). Even without background
subtraction and further analysis, one notices an additition sharp
excitation branch. Theoretical and numerical studies allowed us to
identify this mode with a novel two-magnon bound state
\cite{Schmidiger:11b}. A more thorough data reduction and new
experiments in high magnetic fields are underway, fully justifying
the effort and expense of preparing deuterated crystals.

\subsection{An accidental discovery: an almost perfect $S=1/2$ spin
chain}
\begin{figure}
\begin{center}
\subfigure[]{
\resizebox*{8cm}{!}{\includegraphics[bbllx=50,bblly=330,bburx=560,
  bbury=550,angle=0,clip=]{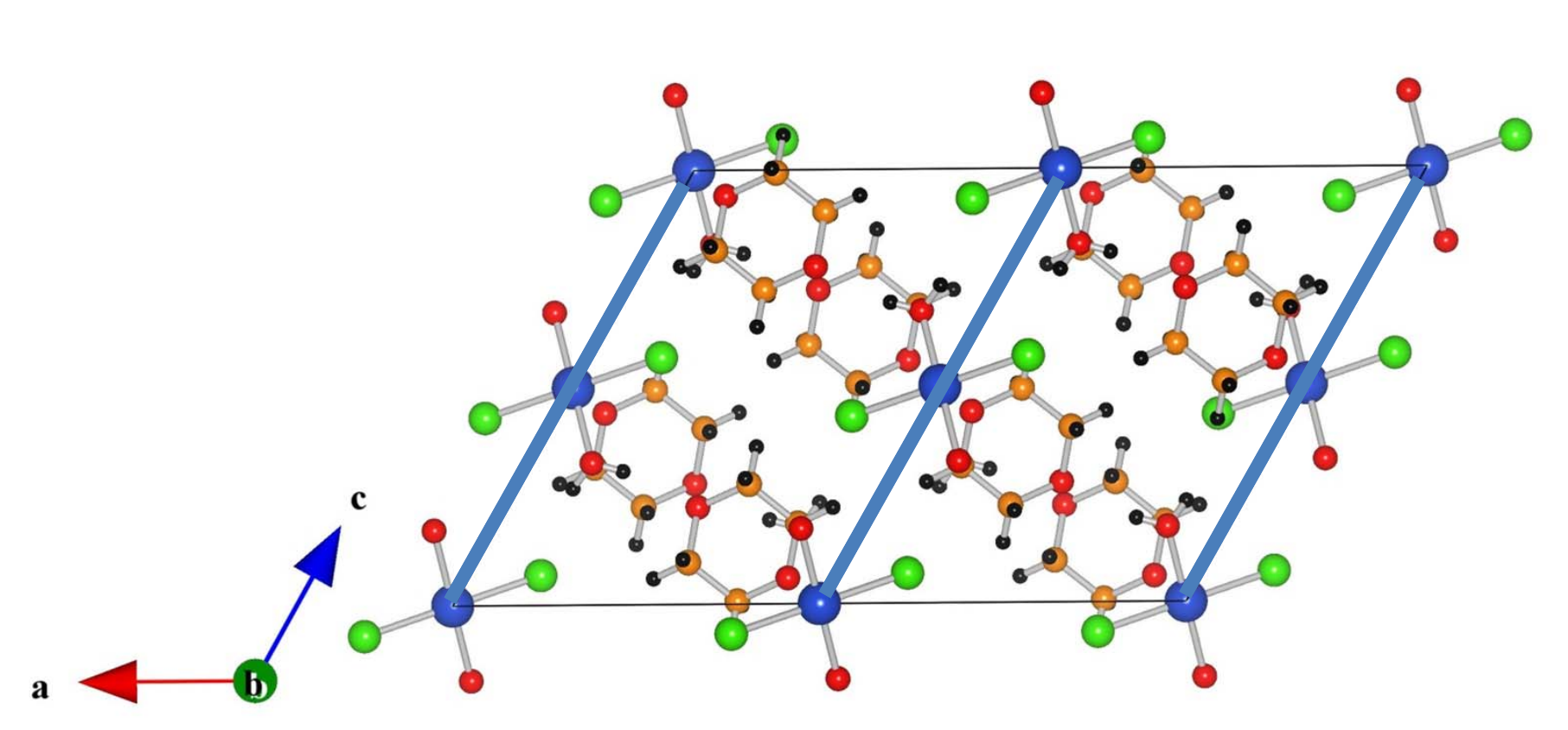}}}%
\subfigure[]{
\resizebox*{5cm}{!}{\includegraphics{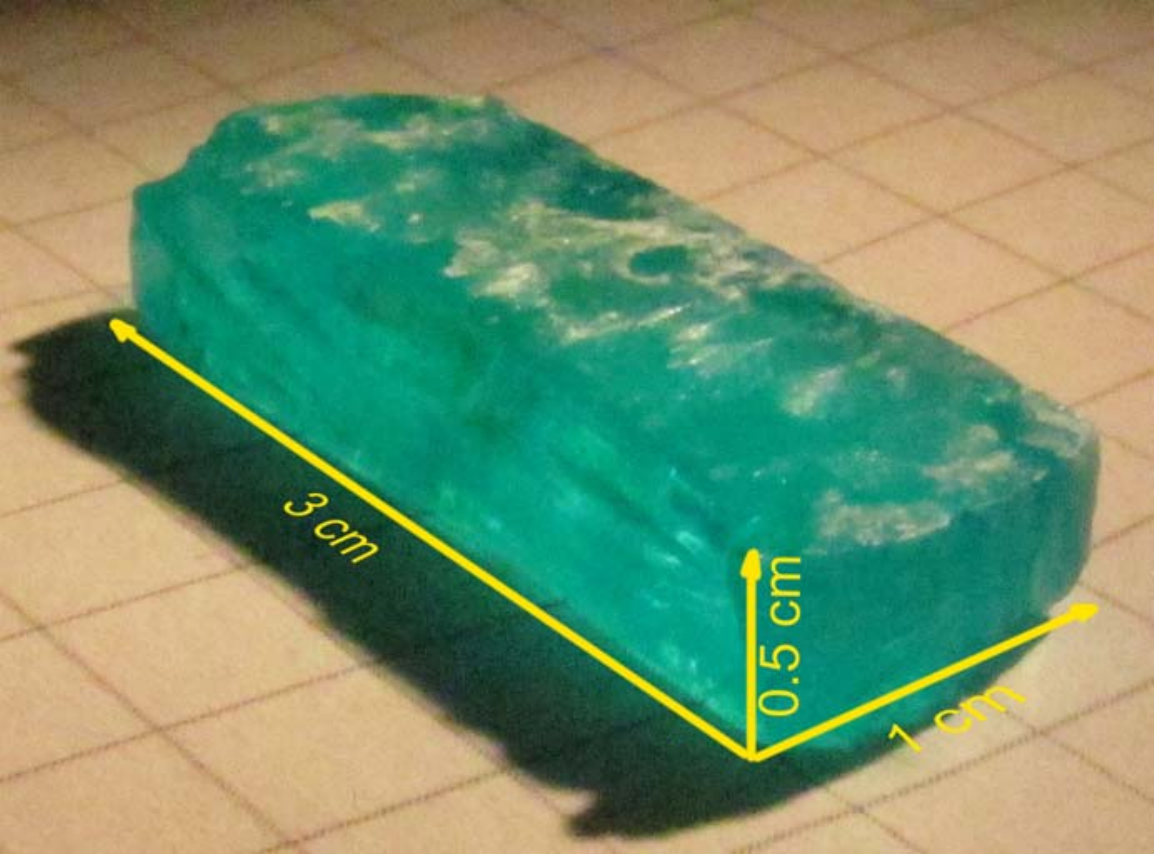}}}%
\caption{(a) View along the crystallographic $b$ axis showing how
Cu$^{2+}$ ions are linked into a one-dimensional chain. The chain
axis is along the crystallographic $c$ direction. The chains are
well separated by 1,4-Dioxane molecules.
Color coding is as follows: Blue:Cu, Green:Cl, Orange:C, Black: H, and Red:O. (b) Shape of a typical 2Dioxane$\cdot$2D$_2$O$\cdot$CuCl$_2$ single crystal grown from aqueous solution with slow evaporation method.}%
\label{dioxane-fig1}
\end{center}
\end{figure}
Solution growth can sometimes produce unexpected results.  It was
reported by Ajiro \emph{et al.} that the material
2Dioxane$\cdot$3CuCl$_2$ realizes a peculiar
ferro-ferro-antiferromagnetic (FFA) Heisenberg $S=1/2$ chain
\cite{Ajiro1994}. With the aim to further investigate this famous
model, at ORNL  we set out to prepare samples following the
procedure  described in \cite{Livermore1982}. The resulting single
crystals seemed to match the description and bulk properties
reported in Ajiro's work. However, once we invested in high-purity
deuterated reagents to grow crystals for neutron scattering, we
suddenly were unable to reproduce the synthesis. The mystery was
solved by studying the original non-deuterated samples with X-ray
diffraction. The data collected on a  Bruker SMART Apex
diffractometer revealed that, in fact, a novel compound,
2Dioxane$\cdot$2H$_2$O$\cdot$CuCl$_2$ was obtained. It appears, that
the necessary water entered the growth solution from the atmosphere
or through the use of low-quality ethanol or methanol solvent. Once
a small amount of heavy water was added to our deuterated reagents,
obtaining large crystals was straightforward.

The slow evaporation growth procedure can be summarized as follows.
3.56 grams of anhydrous CuCl$_2$ is dissolved in 60 ml of deuterated
anhydrous methanol inside a beaker with stirring. 100 ml of
deuterated 1,4-Dioxane and 8 ml of D$_2$O are separately slowly
added consecutively with stirring. The growth beaker is covered
leaving a few pinhole openings. Nicely shaped blue crystals are
harvested after a few weeks. Seeds obtained in this manner can be
suspended on a PTFE filament in the growth solution and further
enlarged. Ultimately, at the laboratory at ETHZ, deuterated crystals
of up to several grams each were successfully grown
(Fig.~\ref{dioxane-fig1}(b)).

Obviously, the new material does not have the unique FFA chain
motif. The crystal structure of
2Dioxane$\cdot$2H$_2$O$\cdot$CuCl$_2$ is monoclinic, with space
group C2/c and lattice constants $a$=17.4298(9) $\AA$, $b$=7.4770(4)
$\AA$, $c$=11.8230(6) $\AA$, and $\beta$=119.4210(10)$^\circ$. It
consists of CuCl$_2$$\cdot$2H$_2$O bi-layers in the $(a,c)$ plane.
These layers are well separated from one another by the 1,4-Dioxane
molecules. The copper ions form simple chains along the
crystallographic $c$ direction as shown in
Fig.~\ref{dioxane-fig1}(a). Intra-chain magnetic coupling is
antiferromagnetic and involves Cu-Cl-H-O-Cu super-exchange pathway.

Crystals of the new material turned out to be unstable in
atmosphere, rapidly losing 1,4-Dioxane to evaporation. Even large
crystals deteriorate in a matter of a few hours. We believe that a
partial decomposition of the sample led to the original
misinterpretation of bulk magnetic measurements. The problem is
completely solved by maintaining the crystals in an atmosphere of
saturated 1,4-Dioxane vapor. Thus, the sample mount used in neutron
scattering experiments is sealed in an aluminium can with a small
amount of deuterated solvent.

\begin{wrapfigure}{l}{0.5\textwidth}
\includegraphics[width=0.48\textwidth]{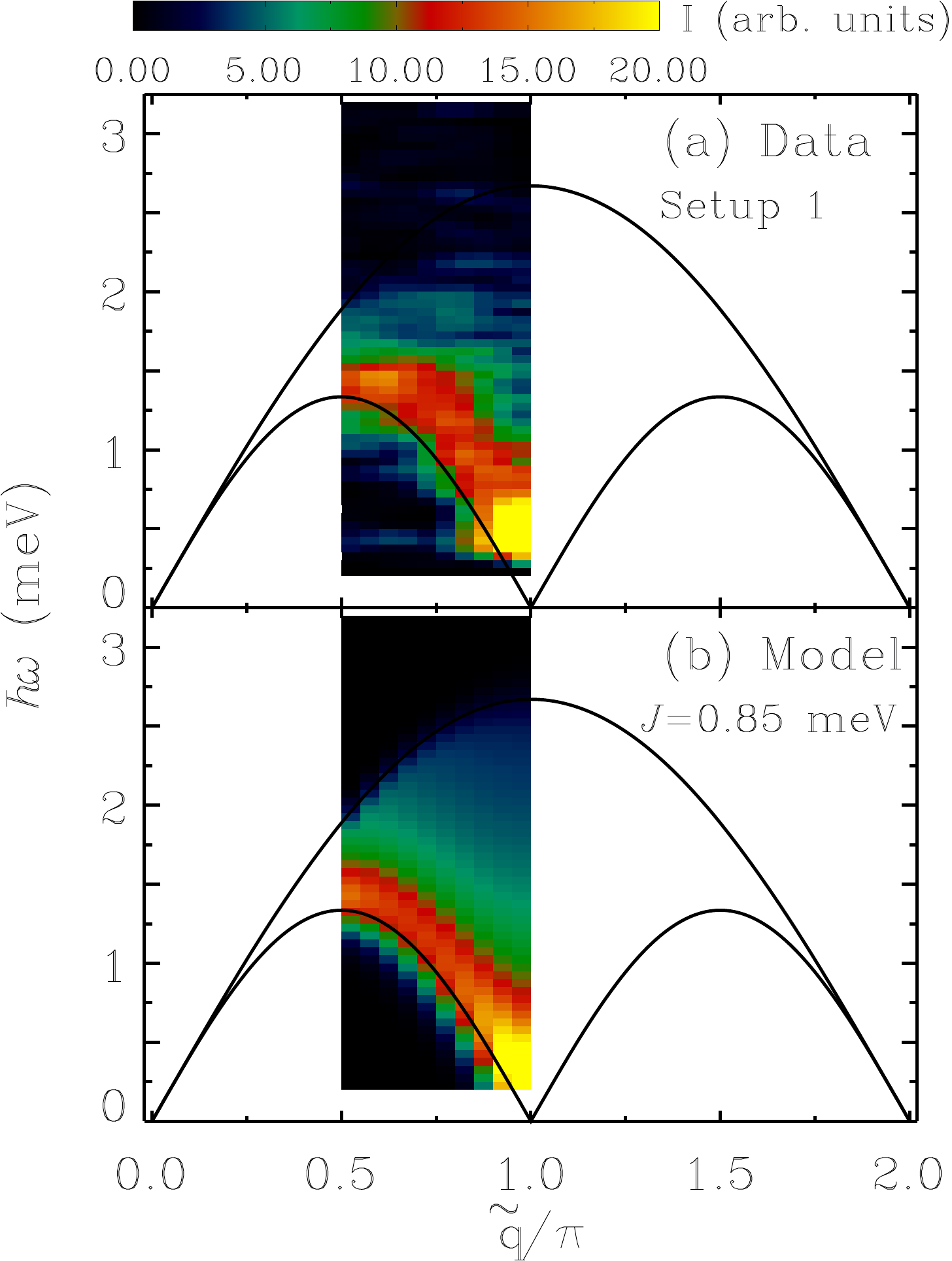}
\caption{(a) False-color contour map of background subtracted
inelastic neutron scattering intensity in
2Dioxane$\cdot$2D$_2$O$\cdot$CuCl$_2$ as a function of transferred
energy $\hbar\omega$ and
 transferred wavevector along the chain $\widetilde{q}$ at \emph{T}=1.6 K \cite{Hong2009}.
 (b) Instrument resolution convolved model calculation of two-spinon continuum M$\ddot{u}$ller
 approximation with \emph{J}=0.85 meV. Solid lines are the predicted lower and upper bounds of the spinon continuum with \emph{J}=0.85 meV.}%
\label{dioxane-fig2}
\end{wrapfigure}

An initial inelastic neutron scattering charachterization of
2Dioxane$\cdot$2H$_2$O$\cdot$CuCl$_2$ was performed at the triple
axes spectrometers TASP at PSI and SPINS at the NIST Center for
Neutron Research (NCNR) \cite{Hong2009}. Fig.~\ref{dioxane-fig2}(a)
shows a false-color contour map of the measured spin excitation
spectrum. These data confirm the simple AF spin chain nature of the
material. In fact, the measured intensity can be attributed to a
multi-particle continuum of so-called spinon excitations, which are
a peculiarity of gapless one-dimensional quantum magnets
\cite{Giamarchibook}. A global fit to the data with the so-called
M$\ddot{u}$ller approximation \cite{Muller1981} of this continuum,
convolved with instrumental resolution, is shown in
Fig.~\ref{dioxane-fig2}(b) The measurements allow us to determine
key energy parameter of this system, namely the intra-chain exchange
constant \emph{J}=0.85 meV \cite{Hong2009}.

As an organic  $S=1/2$ AF chain compound with experimentally
accessible energy scales, 2Dioxane$\cdot$2H$_2$O$\cdot$CuCl$_2$ is
certainly not unique. Among several other representatives, Cu
benzoate \cite{Dender1997} and copper pyrazine dinitrate (CuPzN)
\cite{Stone2003} are perhaps the ones most extensively studied. What
makes the new material special is that, unlike the needle-shaped
CuPzN, it is very easy to grow into extremely large and compact
single crystals. At the same time, unlike Cu-benzoate, it has the
same $g$-tensor for all Cu$^{2+}$ ions, which is critical for
high-field experiments. This makes
2Dioxane$\cdot$2H$_2$O$\cdot$CuCl$_2$ a perfect material for
investigating some very interesting yet subtle features of the
excitation spectrum in applied magnetic fields. In particular, one
expects the dynamic structure factor to obey universal
finite-temperature scaling laws of the Tomonaga-Luttinger liquid
model \cite{Tsvelikbook}. The scaling exponents are predicted to
vary continuously with field \cite{Giamarchibook}. An experimental
verification of these predictions is technically challenging due to
the expectedly very weak neutron scattering intensities. It will
only be possible thanks to the very large deuterated samples
available. These studies are currently underway.

\subsection{A quasi-2D spin liquid}

A significant number of quantum magnets fall into the category of
{\it spin liquids} with a gapped spin excitation spectrum. In these
materials, the only energy scale relevant for low energy physics is
the gap $\Delta$. Excitations from the non-magnetic spin liquid
ground state across the gap can typically be described as dispersive
spin-1 carrying quasiparticles - \emph{magnons}. In the
semiclassical picture, magnons are long-lived. At elevated
temperaturestheir lifetimes are shortened by mutual collisions. Here
dimensionality plays a crucial role. For one-dimensional systems,
reduction of lifetime is the most prominent. Indeed, in $d=1$,
mutual collisions cannot be avoided regardless of the inter-magnon
interaction strength. In the two-dimensional case, the details of
the two-magnon potential become relevant. Magnon decay can also
result from a spatially random \emph{external} potential. Such a
potential can be created, for example, by introducing random
magnetic bonds. How will such disorder affect magnon dynamics?

\begin{figure}[tb]
\begin{center}
\includegraphics[height= 0.4\textwidth]{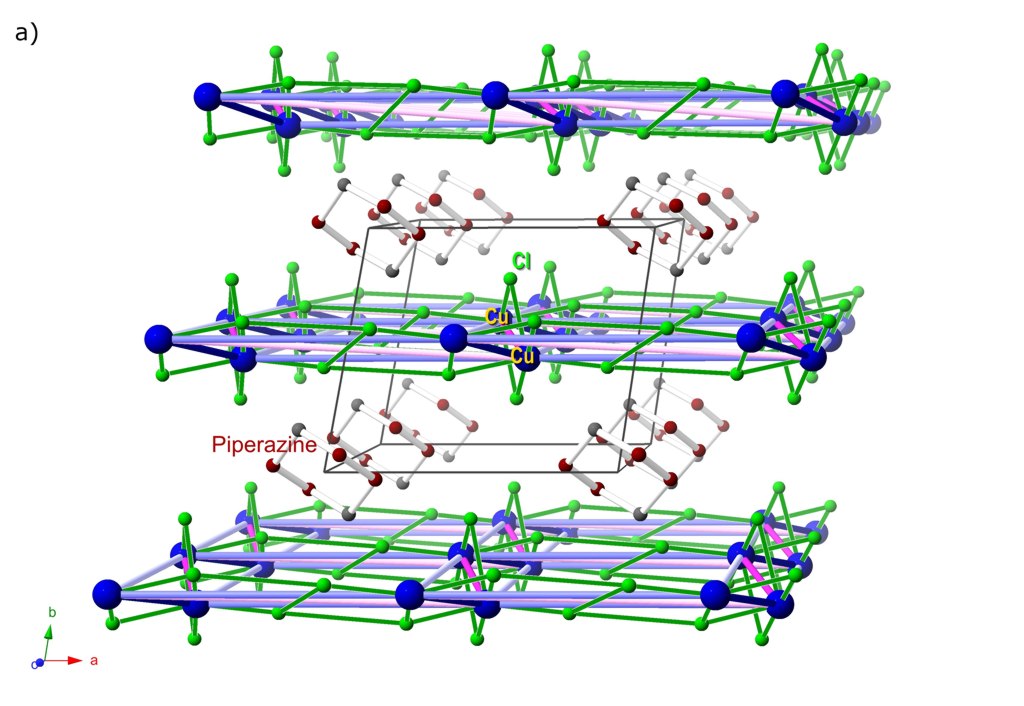}
\includegraphics[height= 0.4\textwidth]{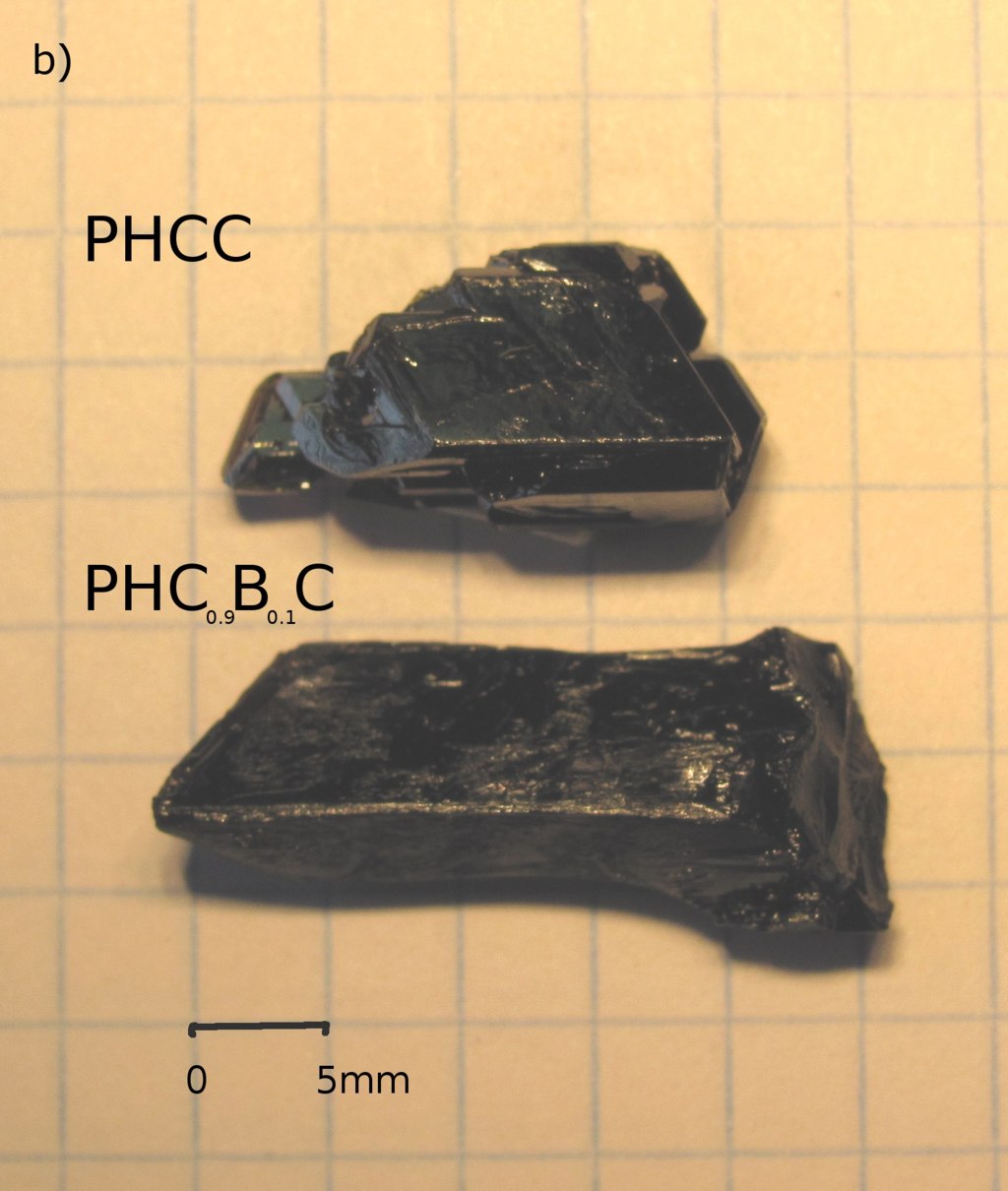}\\
\end{center}
\caption{(a) Crystal structure of PHCC. Antiferromagnetic
superexchange interactions are shown in purple (frustrated) and in
blue (non-frustrated), with bond's strength increasing approximately
with line weight. (b) Large single crystals of pure and 10\% Br
substituted PHCC grown by temperature gradient method.}
\label{phccstruc}
\end{figure}

To tackle this problem experimentally, we chose the
quasi-two-dimesional quantum magnet  piperazinium
hexachlorodicuprate (PHCC). This material crystallizes in the
triclinic space group $P\overline{1}$ ,with a rather large unit cell
($a=6.0836(4)\AA, b=7.0348(5)\AA, c=7.9691(5)\AA,
\alpha=81.287(4)^\circ, \beta=80.053(3)^\circ,
\gamma=68.917(3)^\circ$).  It features $S=1/2$ Cu$^{2+}$ ions in a
complex semi-frustrated network of exchange interactions. The latter
are carried by Cu-Cl-Cl-Cu bonds in the crystallographic
$(ac)$-planes (Fig.\ref{phccstruc}a). The piperazinium molecules
create an effective magnetic insulation layer between these planes.
No inter-plane coupling has been detected to date. The magnetic
properties of pure PHCC have been extensively studied. The value of
the spin gap $\Delta$=1\,meV, as well as all peculiarities of the
spin excitation spectrum, are well
established\cite{Stone2001,Stone2006,Stone2007,Stone2006-Nature}.
 Armed with this knowledge, we set out to introduce disorder in
PHCC by substituting a fraction of the nonmagnetic exchange
interaction-mediating Cl$^-$ ions for the larger ionic radii Br$^-$.
Such a substitution affects the relevant bond angles, and therefore
to {\it locally} modulates the strength of superexchenge
interactions.

The growth of PHCC was first described in Ref.
\cite{marcotrigiano1976} and further characterized in
Refs.\cite{daoud1986,battaglia1988,Stone2001}. To prepare large
single crystals for neutron scattering experiments we adopted the
following procedure. i) Piperazine  (C$_4$D$_10$N$_2$) and CuCl$_2$
powders are separately dissolved in concentrated deuterochloric acid
in a molar ratio 1:4  ii) PHCC powder is formed upon slowly adding
the piperazine solution to the CuCl$_2$ one. iii) Seed crystals are
grown using the temperature gradient method on a hotplate. The
solution is kept at 15 degrees above room temperature. Small
crystals appear on the cylinder walls. iv) The selected seed is
suspended in the growth cylinder. Its surface as well as the
cylinder walls  are periodically cleared of spurious crystals. Using
this method, in four weeks time, we have successfully grown single
crystals of mass up to 2.8g (Fig.\ref{phccstruc}b). Step-like
features tend to form on crystal surfaces but the step's edges and
faces remain strictly parallel to crystal's faces. Growth of
disordered samples (PHCBC) follows exactly the same recipe with
replacement of pure hydrochloric acid with mixture of HCl and HBr in
desired ratio. Our focus is predominantly on small amount of
disorder. Batches of crystals with 0.5\%, 1\%, 2.5\%, 3.5\%, 5\%,
7.5\%, 10\% and 12.5\% nominal Br concentration were synthesized.
Powder and single crystal X-ray measurements confirmed all samples
to be single phase crystals with lattice parameters that increase
monotonically with Br content. Br concentrations  beyond 13\% in
solution seem to lead to phase separation or other instabilities
that inhibit crystal formation. For neutron scattering studies we
replaced the hydrogen rich piperazine with fully deuterated
isotopomer piperazine-d10. The acids were replaced with the
respective deuterated heavy water solutions as well.

\begin{figure}[tb]
\begin{center}
\includegraphics[height= 0.4\textwidth]{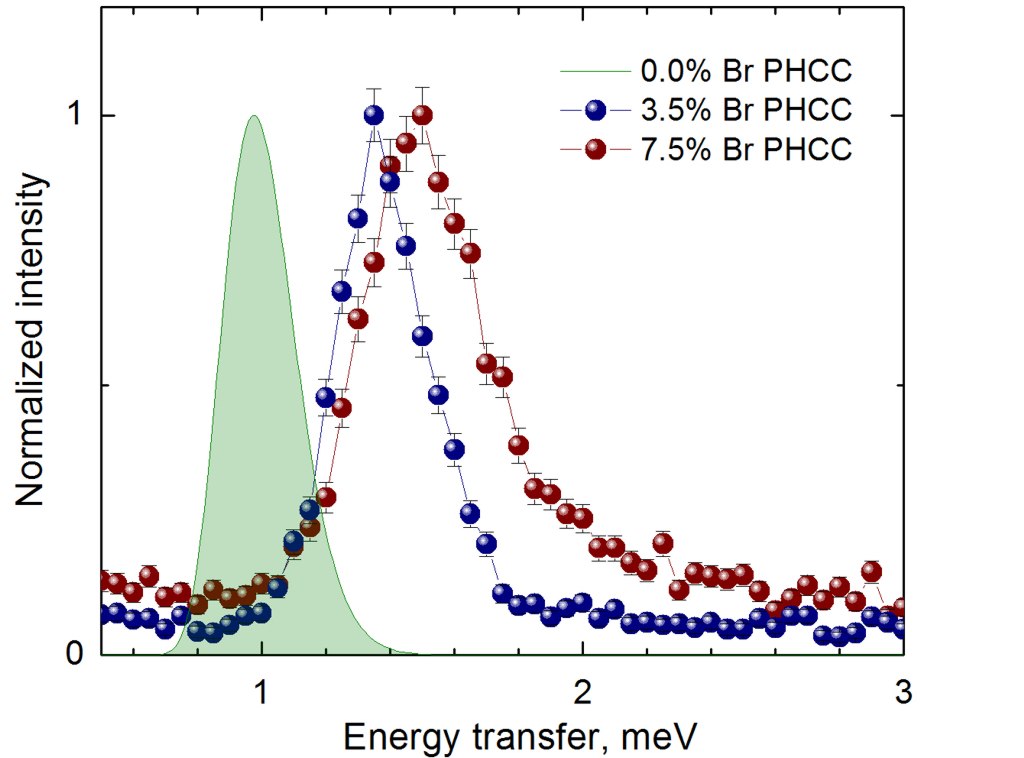}
\includegraphics[height= 0.4\textwidth]{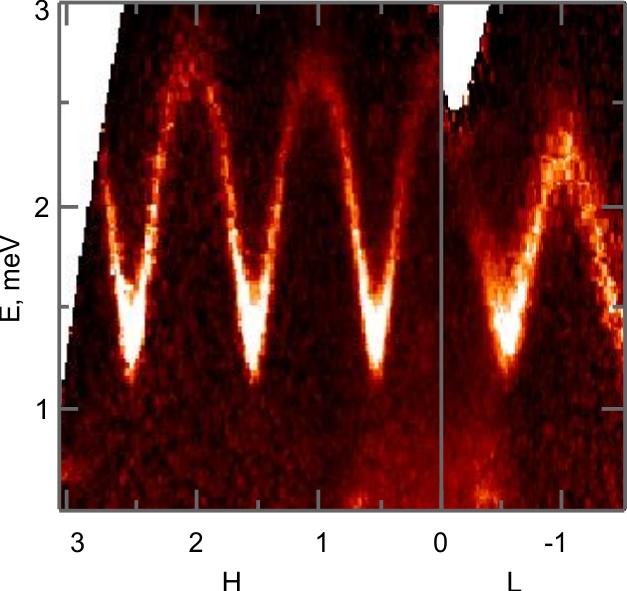}
\end{center}
\caption{\emph{Left:} Effect of disorder on magnon excitations at
the antiferromagnetic zone centre $(0.5, 0.5, -0.5)$, measured at
temperature 1.6~K. The green curve represents instrumental
resolution and is a simulation for disorder-free PHCC based on the
results of \cite{Stone2007}. \emph{Right:} The spin excitation
spectrum measured on CNCS in a 3.5\% Br PHCBC sample, at 1.5~K. }
\label{phccdata}
\end{figure}

The effects of disorder on the magnon excitations are evident from
inelastic neutron scattering experiment. Initial measurements were
performed with triple axis spectrometer TASP ( Fig.\ref{phccdata}).
Compared to what was reported for the disorder-free material
\cite{Stone2007}, the gap energy is increased. In addition, with
increasing disorder, there is a clear broadening of the magnon line,
well beyond the effect of experimental resolution. Our current
efforts are focused on mapping the disorder effects across the
Brillouin zone. In this task, the time-of-flight instrument CNCS at
Oak Ridge National Laboratory has proven to be a most valuable tool,
providing simultaneous access to large range of momentum transfers.
Future plans include the study of the field induced quantum critical
point in disordered PHCC.

\newcommand{\SulPure}{$\text{H}_{\text{8}}\text{C}_{\text{4}}\text{SO}_{\text{2}}\cdot\text{Cu}_{\text{2}}\text{Cl}_{\text{4}}$}
\newcommand{\SulD}{$\text{D}_{\text{8}}\text{C}_{\text{4}}\text{SO}_{\text{2}}\cdot\text{Cu}_{\text{2}}\text{Cl}_{\text{4}}$}
\newcommand{\SulX}{$\text{H}_{\text{8}}\text{C}_{\text{4}}\text{SO}_{\text{2}}\cdot\text{Cu}_{\text{2}}\text{(Cl}_{\text{1-x}}\text{Br}_{\text{x}}\text{)}_{\text{4}}$}
\newcommand{\Sulfolane}{$\text{D}_{\text{8}}\text{C}_{\text{4}}\text{SO}_{\text{2}}$}

\subsection{Frustration}
A broad spectrum of exotic properties is expected to emerge in
quantum spin systems with a {\it geometric frustration} of magnetic
interaction. One such compound is \SulPure (SCC for short). It
crystallizes in a triclinic structure (space group No. 2,
$\text{P}\bar{\text{1}}$) with the lattice constants $a=9.39$~\AA ,
$b=10.76$~\AA , $c=6.62$~\AA , $\alpha = 99.0 ^{\circ}$, $\beta =
95.2^{\circ}$ and $\gamma = 120.7^{\circ}$. The spin $S
=1/2$-carrying Cu$^{+2}$ ions are arranged in ``4-leg spin ladders''
that run along the crystallographic $c$ axis (Fig. \ref{sulstruc}a).
Magnetic interactions are though Cu-Cl-Cl-Cu superexchange pathways.
As usual, the organic ligand acts as a filler ensuring excellent
magnetic one-dimensionality of the Cu$^{2+}$ subsystem. The network
of antiferromagntic interactions is rather complex and has a high
degree of geometric frustration. As a result, for a long time, the
physics of this material was misinterpreted \cite{Fujisawa2003,
Fujisawa2005}.

\begin{figure}
    \label{sulstruc}
    \begin{centering}
\subfigure[][]{\includegraphics[width=0.4\textwidth]{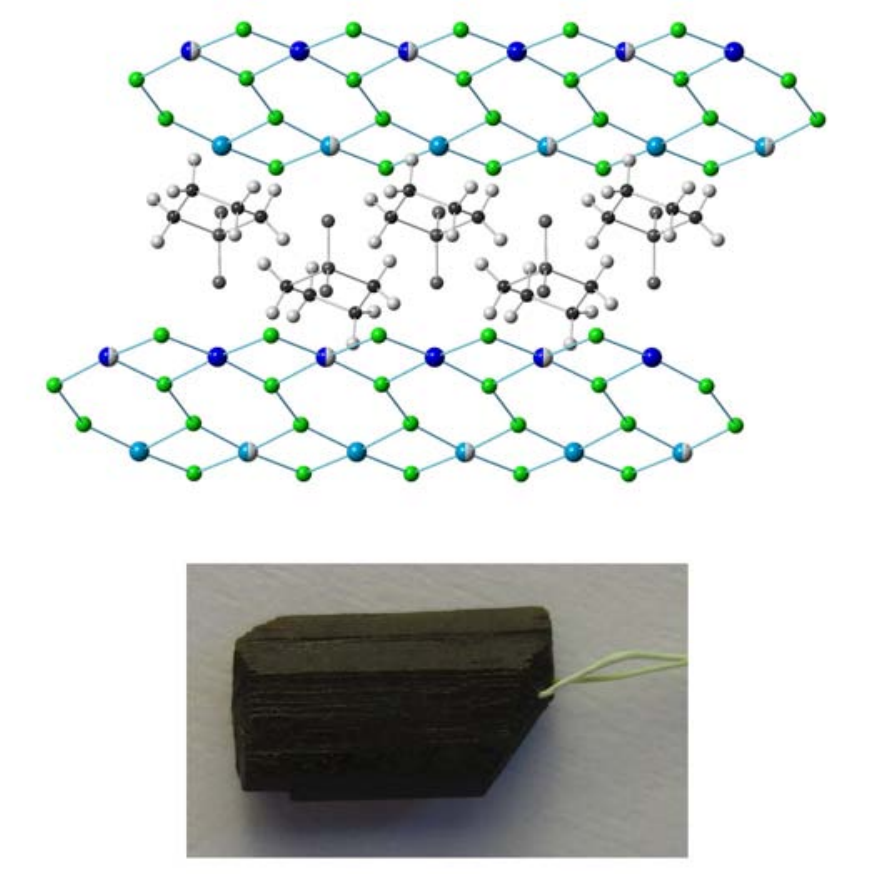}}
\subfigure[][]{\includegraphics[width=0.58\textwidth]{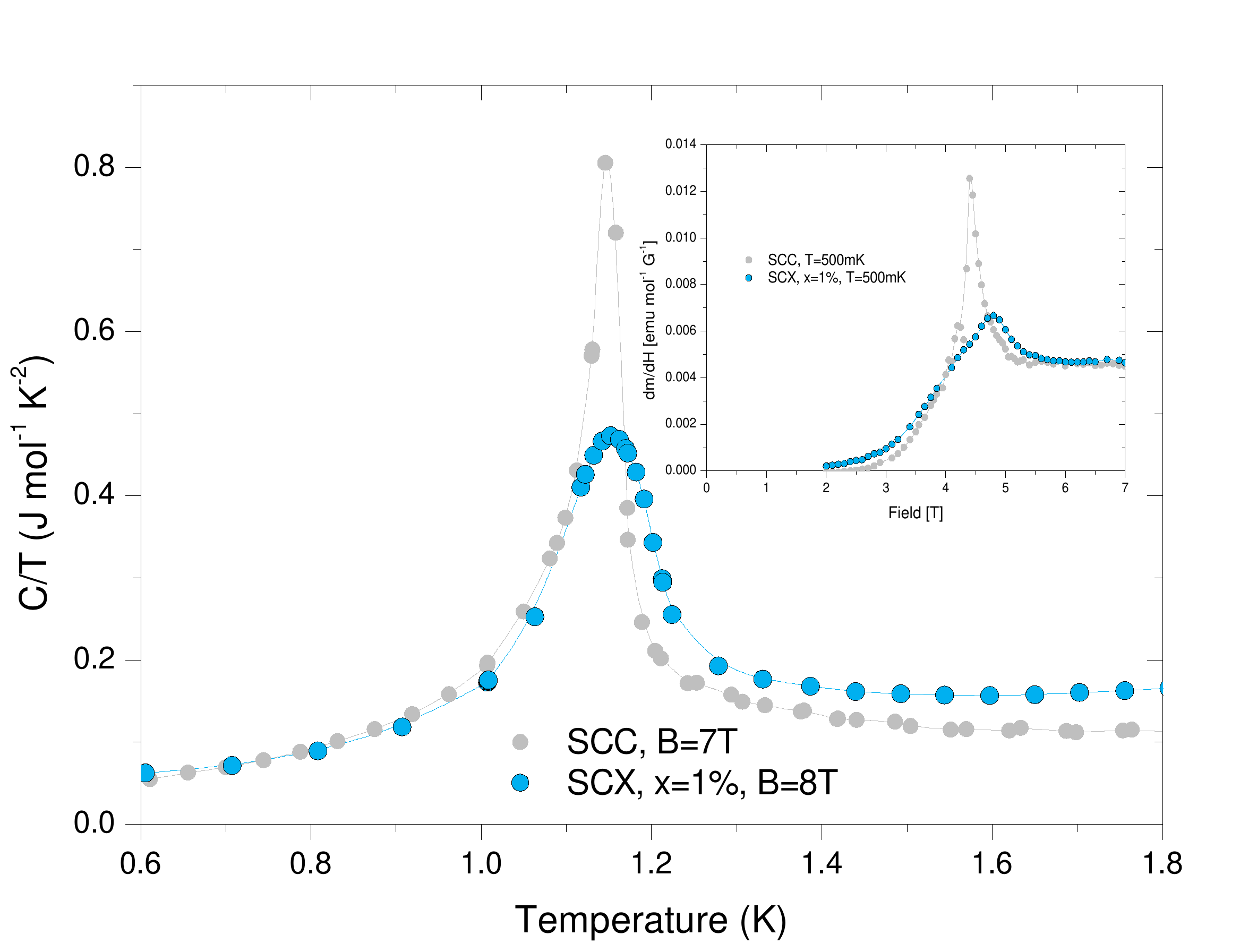}}
\caption{\label{sulstruc}(a) Top: Four-leg spin ladders in \SulX.
The Cu$^{+2}$-ions (blue) interact (light blue lines) via the
Cl$^{-}$ or Br$^{-}$ ions (green). The organic ligand is shown in
gray. Bottom: An $x=0.25\%$ \SulX crystal grown by the temperature
gradient method. (b) Specific heat as a function of temperature
(main panel) and magnetic susceptibility as a function of field
(inset) measured in \SulX for $x=0$ and $x=0.012$
\cite{Wulf2011preprint}.  }
    \end{centering}
\end{figure}

It was inelastic neutron experiments that revealed the true nature
of the magnetic ground state and excitation spectrum. For this
purpose, deuterated single crystals were grown at ORNL by slow
evaporation as described in \cite{Swank1974, Fujisawa2003,
Fujisawa2005}. Copper (II) chloride was dissolved in deuterated
ethanol and deuterated sulfolane (\Sulfolane) was added to the
solution. The solution was then slowly evaporated at 60$^{\circ}$C
until brown crystals were formed on the bottom of the growth beaker.
We were never able to fully optimize this process, and the success
rate remained below 50\%. The resulting crystals were often somewhat
irregular in shape, tending to form as elongated platelets with a
maximal mass of about 200mg. In the end, a sufficient number of
these could be assembled in a ``supersample'' of about 2~g total
mass (Fig.~\ref{Ovi}B). Inelastic neutron data collected at the Disc
Chopper Spectrometer at NIST NCNR (Fig.~\ref{Ovi}A) revealed a small
energy gap of $\Delta\sim 0.5$~meV positioned at an {\it
incommensurate} wave vector $\mathbf{qc}=0.48\pi$ (Fig.~\ref{Ovi}B)
\cite{Garlea2008}.  It was later shown that the the system is best
described in terms of a pair of coupled frustrated 2-leg ladders
\cite{Zheludev2009}, and that incommensurability is a direct
consequence of frustration. Furthermore, elastic neutron scattering
experiments revealed that in an applied magnetic field $H_c=3.75$~T,
the material goes through a quantum phase transition that is a
Bose-Einstein condensation of magnons \cite{Garlea2009}. The
high-field phase is an {\it incommensurate} quantum helimagnet.

\begin{figure}
        \includegraphics[width=1\textwidth]{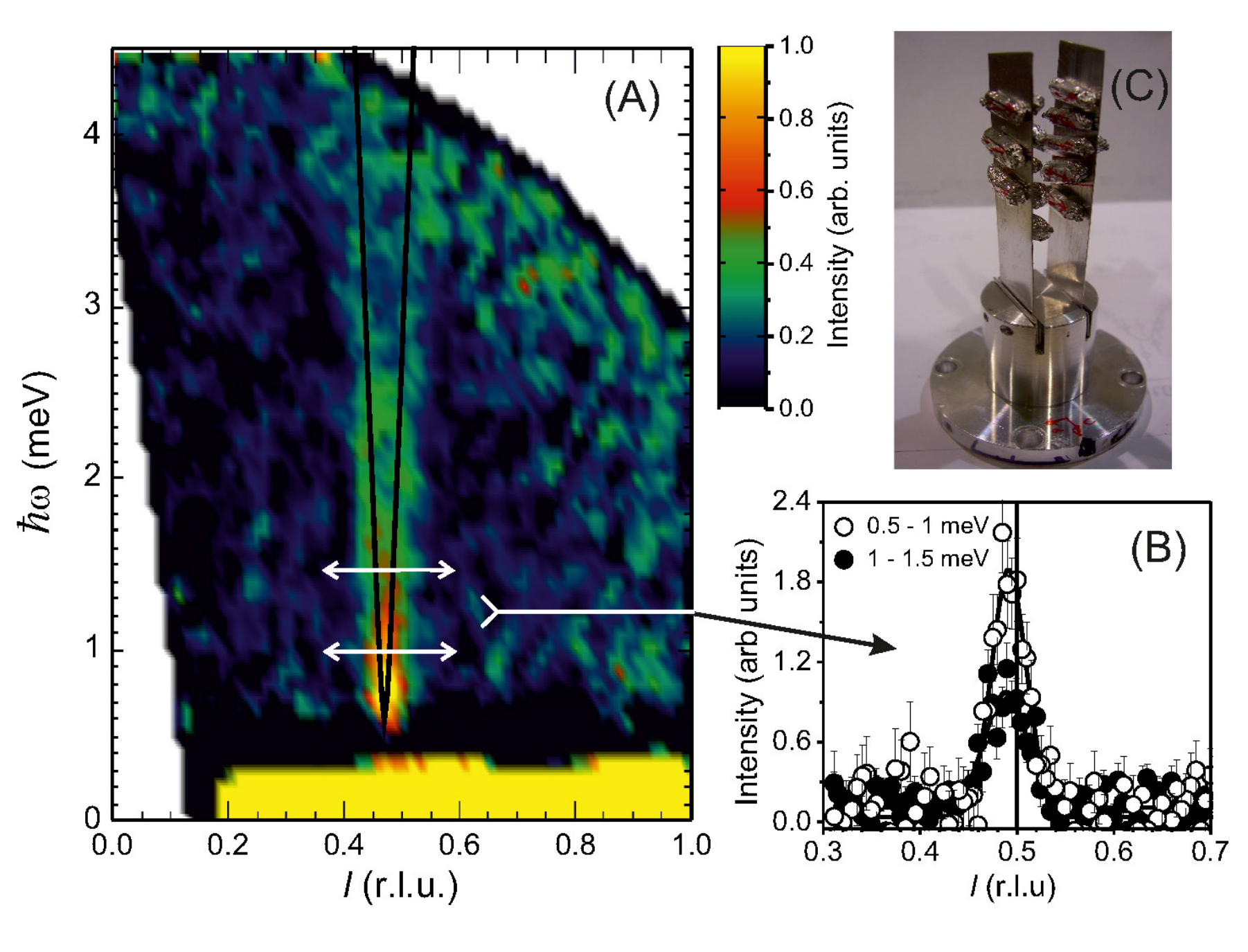}
    \label{Ovi}\vspace{-1cm}
\caption{\label{Ovi}(A) Spin excitation spectrum of \SulD measured
at 70mK using the DCS chopper spectrometer \cite{Garlea2008}.  (B)
Constant-energy cuts along the chain axis reveal the slightly
incommensurate position of the magnon dispersion minimum. (C) Single
crystal assembly consisting of 13 deuterated SCC crystals, used for
this INS study.}
\end{figure}

A very interesting question is how this field-induced quantum phase
transition is affected by bond randomness, of the type studied in
PHCC (see above). We addressed this issue by a fabrication and
systematic study of disordered Br-substituted crystals with the
formula \SulX, or SCX for short. In doing so, we switched to the
temperature gradient growth method. For hot plate temperatures in
the range 45$^{\circ}$C and 160$^{\circ}$ this approach yields much
more reproducible results than slow evaporation. Compact and
perfectly faceted crystals (Fig.~\ref{sulstruc}a) with masses
reaching 1~g were obtained at the ETHZ laboratory. The method works
for Br concentrations $x$ of up to about 12\%. X-ray diffraction
reveals a uniform distribution of substitution sites.

Given considerable investment made in deuterated reagents, it was
quite {\it frustrating} to find that in the disordered material with
Br concentrations as low as 1\%, persistent neutron diffraction
experiments failed to detect any sign of magnetic long range order
in applied fields. Neither incommensurate nor commensurate Bragg
reflections could be observed. This mystery was resolved in
systematic low-temperature bulk measurements, for which the high
quality of the temperature gradient grown samples was crucial. It
turns out that even at very low Br content, field-induced
3-dimensional long-range ordering is indeed suppressed. In place of
a sharp lambda-anomalies in specific heat curves and divergent
magnetic susceptibility seen in the disorder-free compound, in the
Br-substituted material one finds only broad features
(Fig.~\ref{sulstruc}B). We attribute this behavior to a novel
disorder mechanism caused by {\it random geometric frustration} of
magnetic intearctions \cite{Wulf2011preprint}.

\section{Summary}
We would like to re-iterate our main message. In many cases, with
patience and luck, even a quantum physicist can grow large crystals
for neutron scattering studies, and then apply his or her true
expertise to obtain high-quality data and new insight. Drawing the
analogy between crystal growth and the culinary art, one can
rephrase this as a quote from the famous 2007 animation feature
``Ratatouille'': ``Anybody can cook!''

\section{Acknowledgemets}
The authors learned the techniques of sample growth described in
this paper from their colleagues and collaborators, including M. B.
Stone and B. Sales (Oak Ridge National Laboratory), C. L. Broholm
(Johns Hopkins University), F. Xiao and C. Landee (Clark
University), K. Katsumata (previously at RIKEN), H. Manaka
(Kagoshima University), M. Hagiwara (Osaka University), \framebox{P.
Rey} (CEA Grenoble) and many others. All recent measurements on the
TASP spectrometer were facilitated by S. Gvasaliya and M. Mansson of
ETHZ.   We thank Paul Canfield (Ames National Lab) for support and
useful comments and Dr. V. Glazkov (Kapitza Institute, Russian Acad.
Sci.) for his involvement at the early stages of this project. Dr.
T. Hong would like to point out that he was not involved in the part
of the present project related to the DIMPY compound. Single-crystal
X-ray diffraction work by RC was sponsored by the Division of
Chemical Sciences, Geosciences and Biosciences, Office of Basic
Energy Sciences, US Department of Energy. The research at Oak Ridge
National Laboratory's Spallation Neutron Source was sponsored by the
Scientific User Facilities Division, Office of Basic Energy
Sciences, U. S. Department of Energy. One of the authors (AZ), owes
his interest in crystals to his father \framebox{Acad. Prof. I. S.
Zheludev} and to his sister \framebox{Prof. S. I. Zheludeva}, both
of whom were prominent Russian crystallographers. This work is
partially supported by the Swiss National Fund under project
2-77060-11 and through Project 6 of MANEP.

\bibliographystyle{tPHM}

\end{document}